\documentclass[twocolumn ,trackchanges]{aastex62}
\accepted{October, 18 2021}
\shorttitle{321 GHz H$_{2}$O masers in NGC\,4945 and  Circinus galaxy}
\shortauthors{Hagiwara, Horiuchi, Imanishi, \& Edwards}
\newcommand{\kms}{km s$^{-1}\;$}
\newcommand{\kmss}{km s$^{-1}$}
\newcommand{\km}{km s$^{-1}\;$}

\newcommand{\vlsr}{V$_{\rm LSR}$}
\newcommand{\vsys}{V$_{\rm SYS}$}

\newcommand{\msun}{\mbox{M$_{\sun}$}}

\newcommand{\lsun}{\mbox{$L_{\sun}$}}

\newcommand{\ho}{H$_{2}$O$\;$}

\newcommand{\mb}{mJy beam$^{-1}$$\;$}
\newcommand{\mbbb}{mJy beam$^{-1}$}

\newcommand{\jb}{Jy beam$^{-1}$$\;$}
\newcommand{\lwater}{L$_{\rm H_2O}\;$}

\newcommand{\nrest}{$\nu_{\rm {rest}}$}
\begin{document}
\title{Second Epoch ALMA Observations of 321\,GHz Water Maser Emission in NGC\,4945 and the Circinus Galaxy}
\hypersetup{linkcolor=red,citecolor=blue,filecolor=cyan,urlcolor=magenta}

\correspondingauthor{Yoshiaki Hagiwara}
\email{yhagiwara@toyo.jp}

\author[0000-0002-9043-6048]{Yoshiaki Hagiwara}
\affil{Natural Science Laboratory, Toyo University, 5-28-20, Hakusan, Bunkyo-ku, Tokyo 112-8606, Japan}

\author[0000-0002-8395-3557]{Shinji Horiuchi}
\affiliation{CSIRO Space and Astronomy, Canberra Deep Space Communications Complex\\
PO Box 1035, Tuggeranong, ACT 2901, Australia}

\author[0000-0001-6186-8792]{Masatoshi Imanishi}
\affiliation{National Astronomical Observatory of Japan, National Institutes of Natural Sciences (NINS)\\
2-21-1 Osawa, Mitaka, Tokyo 181-8588, Japan}
\affiliation{Department of Astronomy, School of Science, The Graduate University for Advanced Studies, SOKENDAI \\ 
2-21-1 Osawa, Mitaka, Tokyo 181-8588 Japan}
\author[0000-0002-8186-4753]{Philip G. Edwards}
\affiliation{CSIRO Space and Astronomy,\\
 PO Box 76, Epping NSW 1710, Australia}
%
%
%z\author[0000-0002-0786-7307]{Greg J. Schwarz}
%\affil{American Astronomical Society \\
%2000 Florida Ave., NW, Suite 300 \\
%Washington, DC 20009-1231, USA}

%\affiliation{AAS Director of Publishing}
%\affiliation{American Astronomical Society \\
%2000 Florida Ave., NW, Suite 300 \\
%Washington, DC 20009-1231, USA}

%% Note that the \and command from previous versions of AASTeX is now
%% depreciated in this version as it is no longer necessary. AASTeX 
%% automatically takes care of all commas and "and"s between authors names.

%% AASTeX 6.2 has the new \collaboration and \nocollaboration commands to
%% provide the collaboration status of a group of authors. These commands 
%% can be used either before or after the list of corresponding authors. The
%% argument for \collaboration is the collaboration identifier. Authors are
%% encouraged to surround collaboration identifiers with ()s. The 
%% \nocollaboration command takes no argument and exists to indicate that
%% the nearby authors are not part of surrounding collaborations.
%% Mark off the abstract in the ``ab%stract'' environment. 
%Studies of 321GHz H$_2$O maser emission from the two nearby galaxies
% are presented. 
\begin{abstract}
We present the results of second epoch ALMA observations of 321\,GHz
H$_2$O emission toward two nearby active galactic nuclei,
\object{NGC\,4945} and the \object{Circinus galaxy},
together with Tidbinbilla 70-m monitoring of their 22\,GHz \ho masers.
The two epoch ALMA observations show that the strengths of the
321\,GHz emission are variable by a factor of at least a few,
confirming a maser origin.
In the second epoch, 321\,GHz maser emission from \object{NGC\,4945}
was not detected, while for the \object{Circinus galaxy} the flux
density significantly increased and the velocity gradient and
dispersion have been measured.
With the velocity gradient spanning $\sim$110\,\kmss,
we calculate the disk radius
to be $\sim$28\,pc, assuming disk rotation around the nucleus. We also
estimate the dynamical mass within the central 28\,pc to be
4.3\,$\times$\,10$^8$ \msun, which
is significantly larger than the larger scale dynamical mass, suggesting the velocity
gradient does not trace circular motions on that scale.
The overall direction of the velocity gradient and velocity range of
the blueshifted features are largely consistent with those of the
22\,GHz maser emission in a thin disk with smaller radii of
0.1--0.4\,pc and molecular outflows within $\sim$1\,pc from the
central engine of the galaxy, implying that the 321\,GHz masers could trace part of 
the circumnuclear disk or the nuclear outflows.
\end{abstract}

%he long-term variability between the masers in the two different transitions is  for \object{NGC 4945},  while that is not clear for Circinus galaxy.
%Monitoring with the 70-m telescope between 2012 and 2017 reveals that the intensity of the 
%maser in the Circinus galaxy flared to $\sim$ 70 Jy in 2012, the strongest flux density measured in the galaxy,
%and remains \la 10 Jy. The intensity of the maser in NGC\,4945 decreased in this period.
%while the correlation between the maser intensity 
%in the 22 GHz and 321 GHz transitions is not certain from our data.
%\replaced{dfdsfsfdsfdsfds}{new text}
%\explain{explanatory text}
%% Keywords should appear after the \end{abstract} command. 
%% See the online documentation for the full list of available subject
%% keywords and the rules for their use.
%which needs higher angular resolution for resolving distribution of a number of different velocity features.
%% From the front matter, we move on to the body of the paper.
%% Sections are demarcated by \section and \subsection, respectively.
%% Observe the use of the LaTeX \label
%% command after the \subsection to give a symbolic KEY to the
%% subsection for cross-referencing in a \ref command.
%% You can use LaTeX's \ref and \label commands to keep track of
%% cross-references to sections, equations, tables, and figures.
%% That way, if you change the order of any elements, LaTeX will
%% automatically renumber them.
%%
%% We recommend that authors also use the natbib \citep
%% and \citet commands to identify citations.  The citations are
%% tied to the reference list via symbolic KEYs. The KEY corresponds
%% to the KEY in the \bibitem in the reference list below. 
%%%%%%%%%%%%%%%%%%%%%
% NGC4945 Observed from   18-Nov-2016/11:34:01.3   to   18-Nov-2016/12:04:51.3 (UTC)
%%%%%%%%%%%%%%%%%%%%%%
\section{INTRODUCTION} \label{intro}
Luminous \ho masers in the 6$_{16}$--5$_{23}$ transition (rest
frequency, \nrest = 22.23508\,GHz) have been detected in
more than 180 galaxies \footnote{https://safe.nrao.edu/wiki/bin/view/Main/
\\MegamaserCosmologyProjects},
with most of them are detected toward obscured active
galactic nuclei (AGNs) hosting a Seyfert or LINER nucleus
\citep[e.g.,][]{bra94,bra96,bra08,hagi02,hagi03,hagi18,
  linc03a,linc09,hen05,kon06,zhan06,tar11}.
These are known as \ho megamasers.
A fraction of the megamasers are classified as ``nuclear
masers'' as they are associated with AGN-activity in their host galaxy.
These exhibit spectra indicating evidence for emission from sub-parsec--scale
edge-on rotating discs surrounding a super massive black hole
(SMBH) \citep[e.g.,][]{miyo95}, as revealed by Very Long Baseline
Interferometry (VLBI) observations at milliarcsecond (mas) angular
resolution.
 
\ho maser emission occurs at (sub-)millimeter wavelengths in external
galaxies: the pioneering observational study of extragalactic
(sub-)millimeter masers was the discovery of \ho maser in the
3$_{13}-2_{20}$ transition (\nrest\,=\,183.31009\,GHz) toward the
LINER galaxy NGC\,3079 \citep{liz05}, which was later followed by the
detection of a 183\,GHz maser in the nearby AGN NGC\,4945
\citep{liz16}.

\citet{hagi13} conducted a search for \ho masers in the
10$_{29}$$-$9$_{36}$ transition (\nrest = 321.226\,GHz) toward \ho
megamaser galaxies, which resulted in the first extragalactic
detection of a 321\,GHz \ho maser toward the center of the Circinus
galaxy, obtained with the Atacama Large Millimeter/submillimeter Array
(ALMA). The detection of 321\,GHz \ho maser emission toward the center of
NGC\,4945 from this search was reported later
\citep{hagi16,pes16}.

NGC\,4945 is an edge-on type~2 Seyfert galaxy
hosting a Compton-thick nucleus, which is heavily obscured by
foreground material \citep[e.g.,][]{iwa93,puc14}.
The presence of a bright transient X-ray source was reported on the
south-west arm of the galaxy \citep{iso08}.

The Circinus galaxy is one of the closest Compton-thick AGN with a
highly obscured type~2 Seyfert nucleus  \citep[e.g.,][]{are14,yan09}, exhibiting
a kiloparsec-scale one-sided ionization cone at a position angle of
245\degr  \citep{mar94}, which approximately agrees with that of a
nuclear outflow that is optically identified on kiloparsec scales.
Recent ALMA observations of thermalized CO (J=3--2) molecular line emission 
at $\sim$0.3$\arcsec$ angular resolution have traced the
circumnuclear disk ($\sim$70 $\times$ 30\,pc) and spiral arms within
$\sim$200\,pc of the nucleus \citep{izu18}.

Prominent 22\,GHz \ho masers in the center of the galaxies
\citep{dos79,whi86,linc97b} were studied at milliarcsec (mas) spatial
resolution using VLBI.
In the Circinus galaxy, these observations revealed the presence of an
edge-on disk-like linear structure surrounding what is a most likely a
SMBH.
It is claimed that some fraction of the maser emission in the Circinus
galaxy traces molecular outflows emanating within $\sim$1\,pc from the
nucleus of the galaxy \citep{linc03b}.
In NGC\,4945, an edge-on linear structure extending over
0.7\,pc at a position angle of 45\degr, closer to that of
the galactic disk of the galaxy, was revealed \citep{linc97b,linc03b}.

Currently, the only other extragalactic sub-millimeter \ho line emission
detected is the 5$_{15}$$-$4$_{22}$ transition
(\nrest\,=\,325.15292\,GHz) toward the ultraluminous infrared galaxy
(ULIRG), Arp\,220 \citep{cer06, gala16, kon17}, although thermalized
molecular gas emission is favoured as the origin due to lack of
intensity variability and the broad line-width of the emission.
Very recently, luminous 183\,GHz \ho maser emission (\lwater $\sim$
10$^4$\lsun) has been detected toward the ULIRG, IRAS 19254$-$7245
(the ``Superantennae galaxy'') at $z$=0.0617 \citep{ima21}, providing the
possibility of studying dense molecular gas around a SMBH beyond the
local universe.

Similar to the 22\,GHz nuclear megamasers, sub-millimeter \ho masers
can probe the nuclear kinematics of AGN or dense molecular gas
structures such as a rotating disk surrounding a SMBH, which provides
new insights into the nuclear region of nearby AGN.

In this paper, we present new results
from second epoch observations of the 321\,GHz \ho masers in
NGC\,4945 and the Circinus galaxy, conducted with ALMA at higher angular
resolution.
The optical velocity definition is adopted throughout,
and the velocities are calculated with respect to the Local Standard
of Rest (LSR).

\section{OBSERVATIONS AND DATA REDUCTION} \label{obs}
\subsection{ALMA Cycle 4 observations} \label{obs1}
We conducted Cycle 4 ALMA observations of \ho masers in the
$10_{29}-9_{36}$ transition (\nrest\,=\,321.226\,GHz) towards
\object{NGC\,4945} and the \object{Circinus galaxy} for project
2016.1.00150.S (PI: Y.~Hagiwara) in November 2016 and May 2017.
These band~7
spectral-line observations lasted for about 30--35 minutes
(approximately 8--10 minutes on-source), with 45--47 antennas,
depending on the target source.  The observations were made with a
single spectral window (1875\,MHz bandwidth) and one 
polarization, divided into 7680 spectral channels, yielding spectral
resolutions of 244.1\,kHz or 0.228\,\kms at each observed frequency and
a total velocity coverage of $\sim$1750\,\kms in order to cover
Doppler-shifted emission within $\pm$900\,\kms of the
galaxy's systemic velocity. The spectral windows were centered near
the systemic velocity. After the channel smoothing to 3840 spectral
points, the resulting spectral resolution is 488.3\,kHz or 0.458\,\kms at
the observed center frequency to match that
of the 22\,GHz \ho maser observations at Tidbinbilla (see Section \ref{obs2}).
The observations were made at $\sim$0.3$\arcsec$ angular resolution. A
summary of the observations is listed in Table~$\ref{tab1}$.

%calibration
Data calibration was performed using the Common Astronomy Software
Applications (CASA)  \footnote{\url{https://casa.nrao.edu}}\citep{mcm07}.  Flux density and bandpass
calibration were performed using J1427$-$4206 and phase calibration
was made using J1326$-$5256 for the observations of NGC\,4945.  The
flux density, bandpass, and phase calibrators were J1617$-$5848, J1427$-$4206,
and J1424$-$6807, respectively, for the Circinus galaxy. We adopted a
flux density error of 10 percent, as stated in the capabilities
of Cycle~4 Band~7 observations.

%imaging
Imaging of the calibrated data was carried out with CASA and the
NRAO Astronomical Image Processing Software (AIPS)\citep{van96}.  After phase
and amplitude calibrations, the 321\,GHz continuum emission of
NGC\.4945 and the Circinus galaxy was subtracted from the spectral line
visibilities using line-free channels prior to CLEAN deconvolution of
line emission. The line emission from each galaxy was separated out
from the continuum emission. Imaging was performed using CASA with
Briggs weighting (robust\,=\,0.5). The synthesized beam sizes and the
beam position angles used in the CLEAN process and the resultant rms
noise levels in the spectral line images are listed in
Table~$\ref{tab1}$.

\subsection{Tidbinbilla Observations}\label{obs2}
\ho maser spectra of the galaxies in the $6_{16}-5_{23}$
transition (\nrest=22.23508\,GHz) were obtained using  NASA Deep Space
Network 70-m antenna at Tidbinbilla, {DSS-43}. 
Observations of the masers from June 2012 to April 2014 were conducted using a K-band receiver and ATNF correlator 
with a 64 MHz bandwidth, 2048 spectral channels, and dual circular polarization at a spectral resolution of 31.25\,kHz or
0.42\,\kmss \citep{hagi13,hagi16}, and observations from April 2016 to May 2017 were conducted using a new K-band receiver and SAO Spectrometer with a 1 GHz bandwidth, 32000 spectral channels, and dual circular polarization \citep{kui19} at a spectral resolution of 31.25 kHz or 0.42\,\kmss. We used a velocity resolution of
better than $\sim$1\,\km to avoid blending of different velocity features.
We adopted a telescope sensitivity of 2.6\,Jy\, K$^{-1}$
to convert antenna temperature to flux density with a conservatively
estimated uncertainty in the flux density of 50\%, as stated in the Tidbinbilla 70\,m radio telescope guide\footnote{https://www.atnf.csiro.au/observers/docs/tid\_obs\_guide \\
/tid\_obs\_guide.html} or in \citet{kui19}.
Data reduction, including baseline calibration, averaging the two circular polarizations, and flagging 
 bad scans when required were carried out using GBTIDL\footnote{\url{https://gbtidl.nrao.edu}.
 Analysis of the data obtained om ATNF Correlator was made using ASAP\footnote{\url{https://svn.atnf.csiro.au/trac/asap}}.}
 
All velocities in the final plots are then converted relative to the Local Standard of Rest using the optical definition.
 Rms noise levels (1\,$\sigma$) estimated from each spectrum were $\sim$50--300\,mJy for NGC\,4945 and $\sim$40--90\,mJy for the Circinus galaxy in each 0.42\,\kms channel.  
A summary of the observations is given in Table~$\ref{tab2}$.

\section{OBSERVATIONAL RESULTS} \label{result}
\begin{figure*}[t!]
\epsscale{1.15}
\plottwo{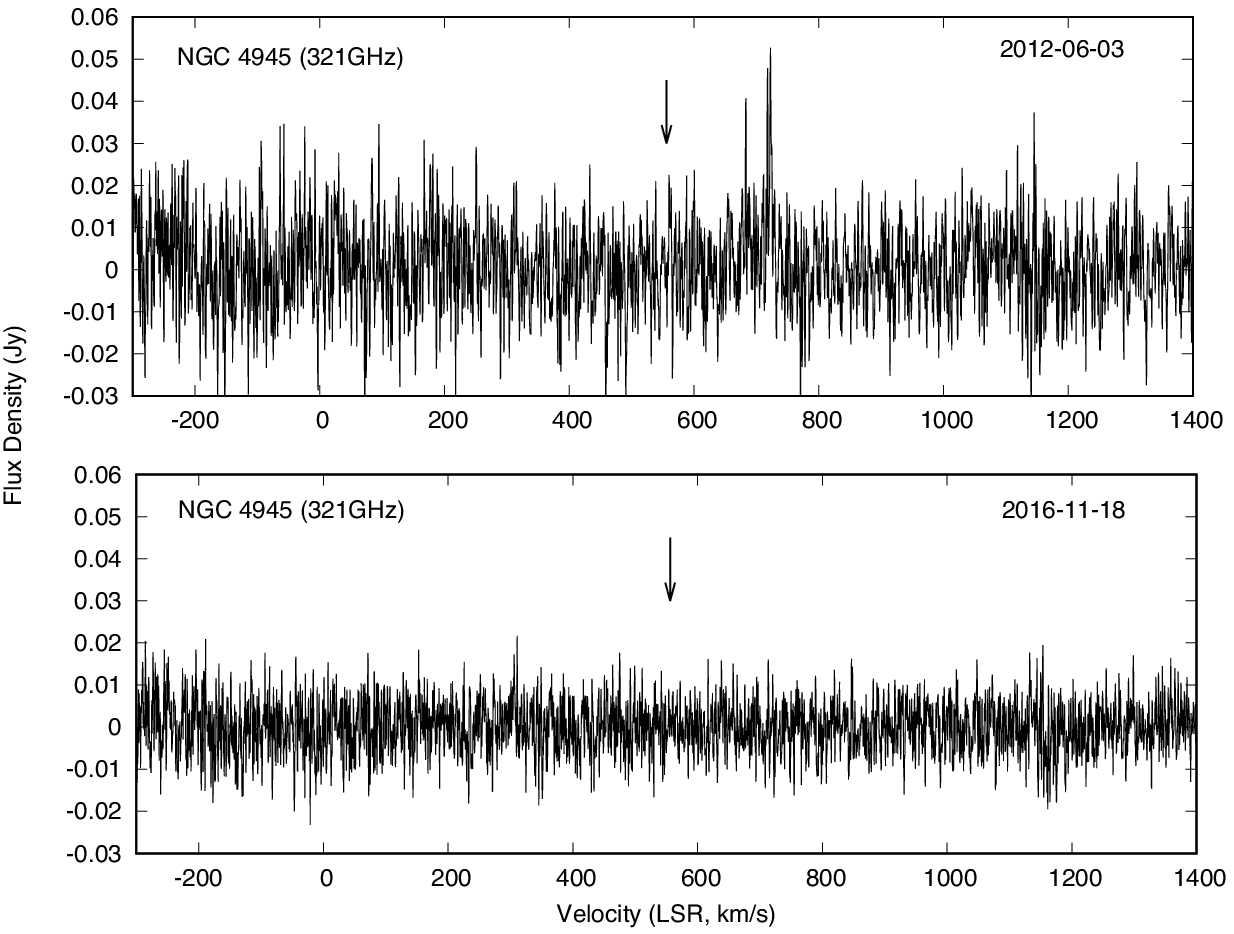}{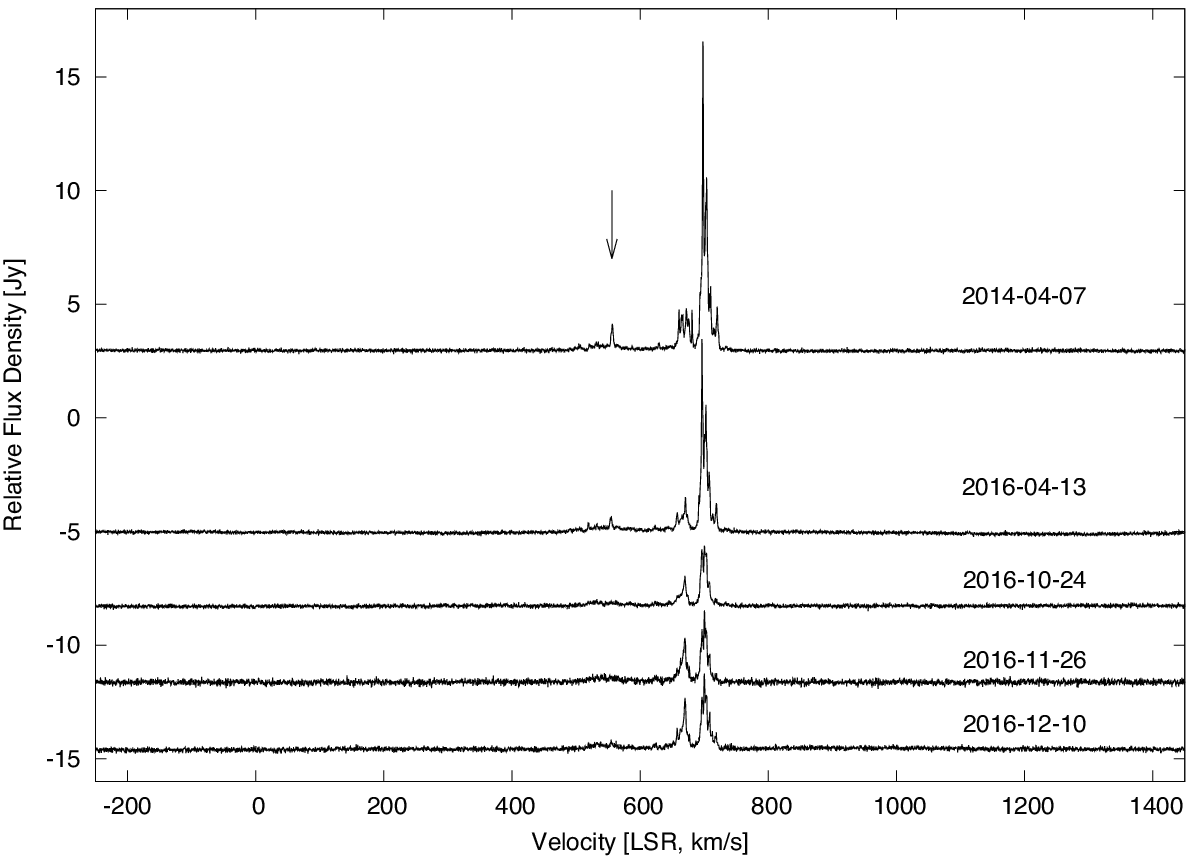}
\caption{Spectra of \ho maser emission toward the nucleus of
  NGC\,4945. The LSR systemic velocity of NGC\,4945, 556\,\kms is
  denoted by the vertical arrow. $Left$: Spectra of 321\,GHz \ho maser for
  two epochs, obtained with ALMA. The velocity resolution is 0.458
  \kmss. $Right$: Spectra of 22\,GHz \ho maser, obtained with the NASA Deep
  Space Network 70 m antenna at Tidbinbilla, {DSS-43},  for five epochs from 2014
  April to 2016 December.  The vertical axis shows relative flux
  density in Jy. The amplitude scale of each 22\,GHz spectrum is
  displayed with an offsets for display purposes. The velocity
  resolution is 0.42\,\kmss. The 1\,$\sigma$ noise levels are $\sim$28--55\,mJy
  in a 0.42\,\kms channel. \label{fign4945}}
\end{figure*}

Following the approach used for the first detections in 2012,
observations of the 321\,GHz \ho masers toward
the center of NGC\,4945 and the Circinus galaxy were centered on
the phase-tracking center of the galaxy in the velocity range of
\vlsr=$-$319.5 -- 1436.5\,\kms for NGC\,4945 and \vlsr=$-$439.9 -- 1314.1\,\kms
for the Circinus galaxy, covering the total velocity range of the
known 22\,GHz \ho maser emission in the galaxies
\citep{linc97b, bra03, linc03a}.
{No search was made for 321\,GHz maser emission beyond these velocity on this occasion.}
Figure~\ref{fign4945} (left) compares the two epoch spectra of
the 321\,GHz \ho maser toward the center of NGC\,4945. No maser
emission is detected in the second epoch (2016 November) as the strength
of the maser has decreased below the detection limit (3\,$\sigma$) of
38\,mJy.  Also, a counterpart to the highly redshifted velocity feature at
\vlsr= 1138.6\,\kms in the 183\,GHz \ho maser spectrum \citep{liz16}
was not found in the second epoch spectrum.
Figure~\ref{fign4945} (right) displays
multi-epoch spectra of the 22\,GHz \ho maser emission that were taken
from 2014 April to 2016 December with the Tidbinbilla 70-m telescope, {DSS-43}, showing
a definite decrease of the maser intensity from mid-2014, at least,
until late 2016.

\begin{figure*}[ht!]
\epsscale{0.8}
%\plottwo{f2a.eps}{f2b.eps}
\plotone{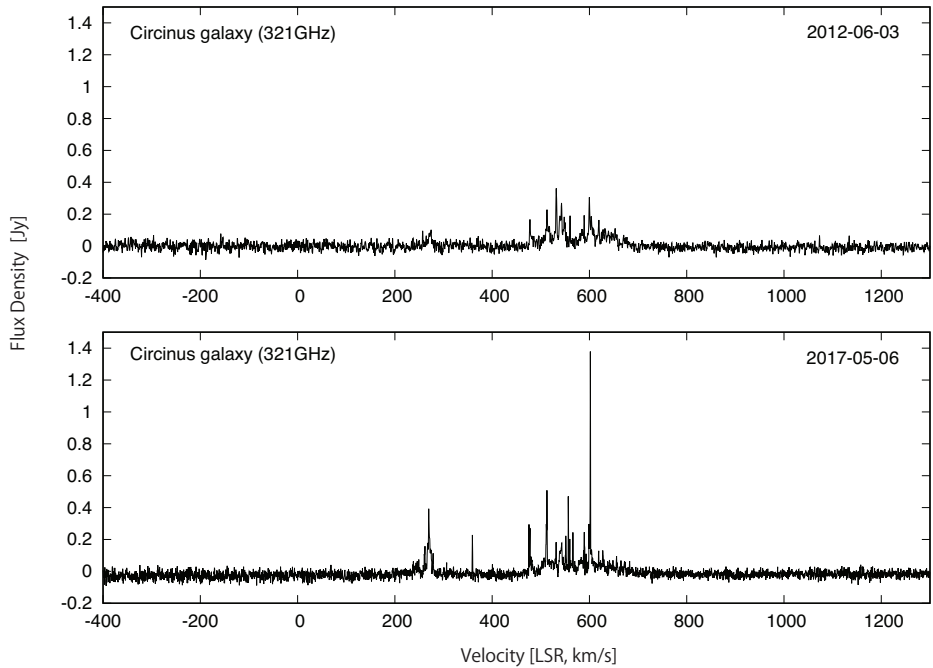}
\caption{Comparison of 321\,GHz \ho maser
  spectra between 2012 June and 2017 May, obtained with ALMA. The
  velocity resolution is 0.458\,\kmss.  \label{circinus321}}
 \end{figure*}

\begin{figure*}[ht!]
\epsscale{0.8}
\plotone{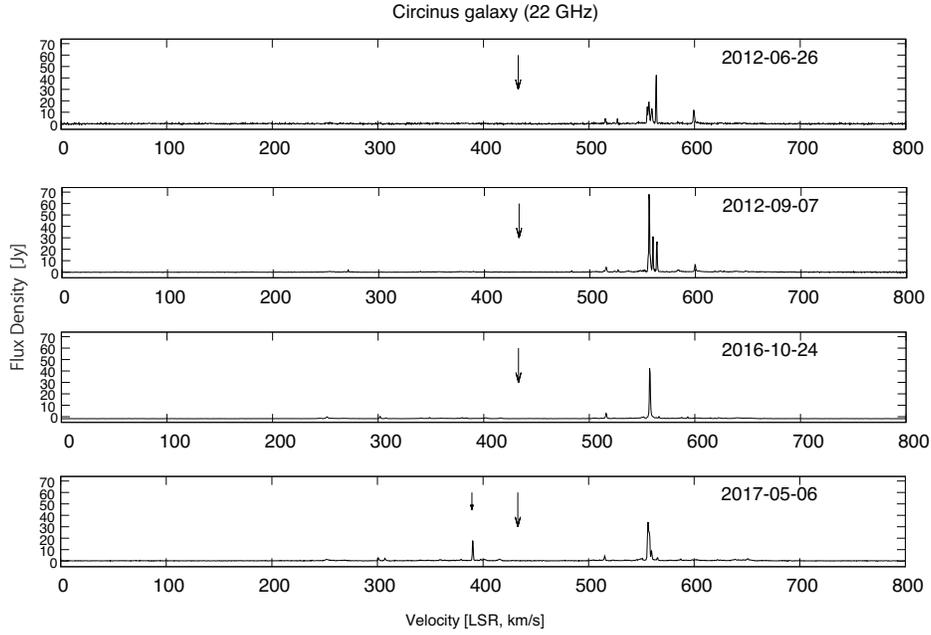}
\caption{Spectra of 22\,GHz \ho
  maser for four epochs, obtained at Tidbinbilla from 2012 June to
  2017 May.  The LSR systemic velocity of the galaxy, 433\,\kms
  is indicated by a longer arrow in each epoch and a shorter
  arrow indicates blueshifted features centered at \vlsr\,=\,389.5\,\kmss. \label{circinus22}}
\end{figure*}

Figure~\ref{circinus321}  shows spectra of the 321\,GHz \ho maser emission
toward the center of the Circinus galaxy in 2012 June and 2017 May, and
Figure~\ref{circinus22} displays four-epoch spectra of 22\,GHz
\ho maser emission taken at Tidbinbilla from 2012 June to 2017 May.  It should
be noted that the second epoch spectrum at 321\,GHz  and the
last epoch spectrum at 22\,GHz were obtained on the same
date, making it possible to precisely compare velocities of the maser
features between the 22\,GHz and 321\,GHz maser spectra.

In the Circinus
galaxy, the most prominent \ho maser emission is detected with a
peak flux density of 1.56\,Jy at \vlsr\,=\,601.6\,\kms with a signal-to-noise
ratio (SNR) of $>$80 and a 1\,$\sigma$ rms level of $\sim$19\,mJy
per spectral channel (Figure~\ref{circinus22}). The line-widths of
the detected Doppler-shifted features may be narrower ($<$0.445\,\kmss)
than one spectral channel,  and narrower than the 22\,GHz \ho maser features
of $\sim$0.9\,\kms reported previously \citep{naka95,linc97a,bra03,mac05}.
In the second epoch observation
of the 321\,GHz \ho maser, blueshifted features spanning from
\vlsr=358--360\,\kms are evident with a peak at \vlsr= 359.29\,\kmss,
Doppler-shifted $\sim$\,$-$74\,\kms from the systemic velocity of
\vlsr\,=\,433\,\kms \citep{dev91,bra03}, with a peak flux density
of 198\,mJy. 
Blueshifted features between \vlsr\,=\,200 and 400 \kms were also found
in the 22\,GHz maser spectrum on the same day. The most blueshifted and
redshifted velocity features peak at \vlsr\,=\,260.81\,\kms and
\vlsr\,=\,627.25\,\kmss, respectively. The total integrated intensity
estimated from the spectra of the 321\,GHz \ho features detected at
SNR $>$5 between \vlsr = 200 and 800\,\kms is 11.1$\pm$1.1\,Jy\,\kmss.

\begin{figure*}[ht!]
\epsscale{1.16}
\plottwo{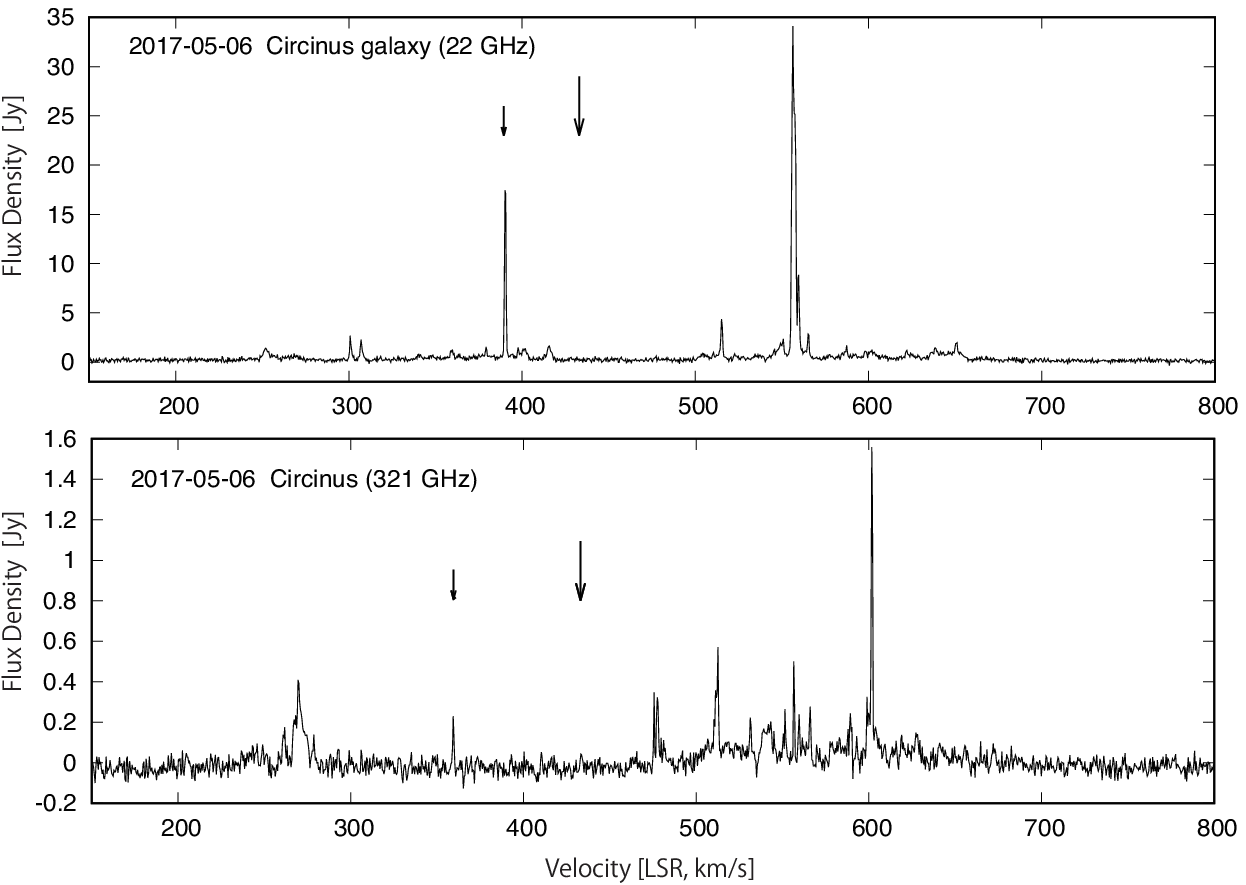}{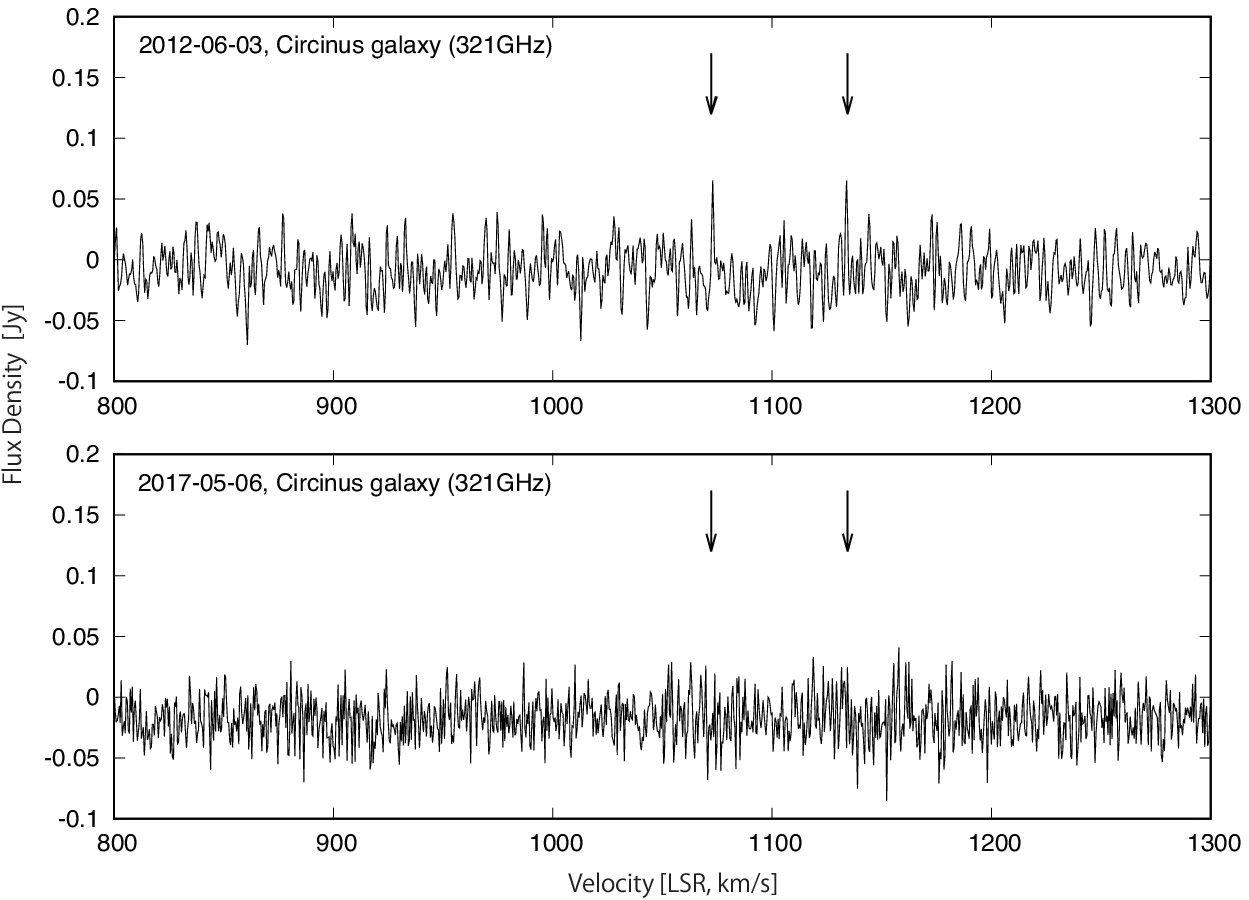}
\caption{$Left$: Comparison of 22\,GHz ($upper$) and 321\,GHz ($lower$)
  \ho maser spectra toward the nucleus of Circinus galaxy zoomed in
  the velocity range of \vlsr\,=\,150--800\,\kms in
  Figure~\ref{circinus321}. Note that the both spectra were taken on 2017
  May 6. The longer arrow in each panel shows the systemic velocity, and
  \vlsr\,=\,389.49\,\kms ($upper$) and \vlsr\,=\,359.29\,\kms ($lower$)
  features are marked by a shorter arrow.  $Right$: Highly
  Doppler-shifted (high-velocity) 321\,GHz maser spectra toward the
  nucleus of the Circinus galaxy in the velocity range of
  \vlsr\,=\,800--1300\,\kms at epoch~1 ($upper$) and epoch~2 ($lower$).
  The channel spacing is 0.458\,\kmss.
  In the epoch~1 spectrum, high-velocity tentative
  features were suggested at \vlsr\,=\,1069\,\kms and 1129\,\kms (shown by
  arrows)\citep{hagi13}. However, in the epoch~2 spectrum no emission
  was observed at these frequencies.  \label{highv}}
\end{figure*}

Figure~\ref{highv} (left) shows the spectrum of the 321\,GHz maser emission
spanning from \vlsr\,=\,180$-$820 \kms in the Circinus galaxy.
Figure~\ref{highv} (right) compares the two epoch spectra at
321\,GHz in the highly redshifted velocity range.
Signs of marginal features were suggested at \vlsr\,=\,1069\,\kms
and \vlsr\,=\,1138.6\,\kms in \citet{hagi13}, however these features
were not identified in the second epoch spectrum. As argued by \citet{pes16},
it is most likely that this portion of the band is impacted by higher atmospheric absorption
resulting in a somewhat higher noise level. 
%1069 \kms feature (redshifts up to $\sim$ 635 \kmss) 

\begin{figure*}[ht!]
\epsscale{1.16}
\plottwo{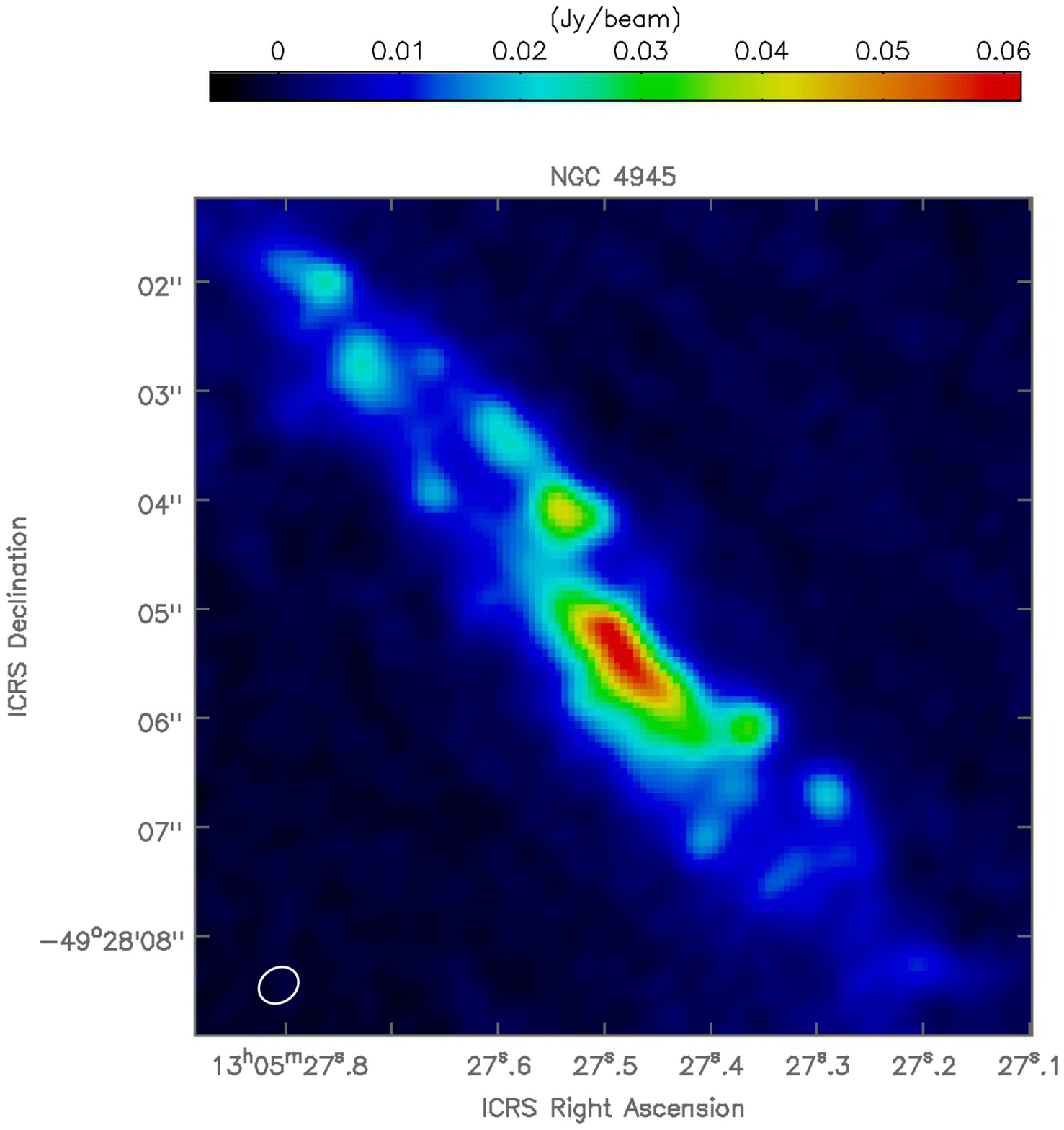}{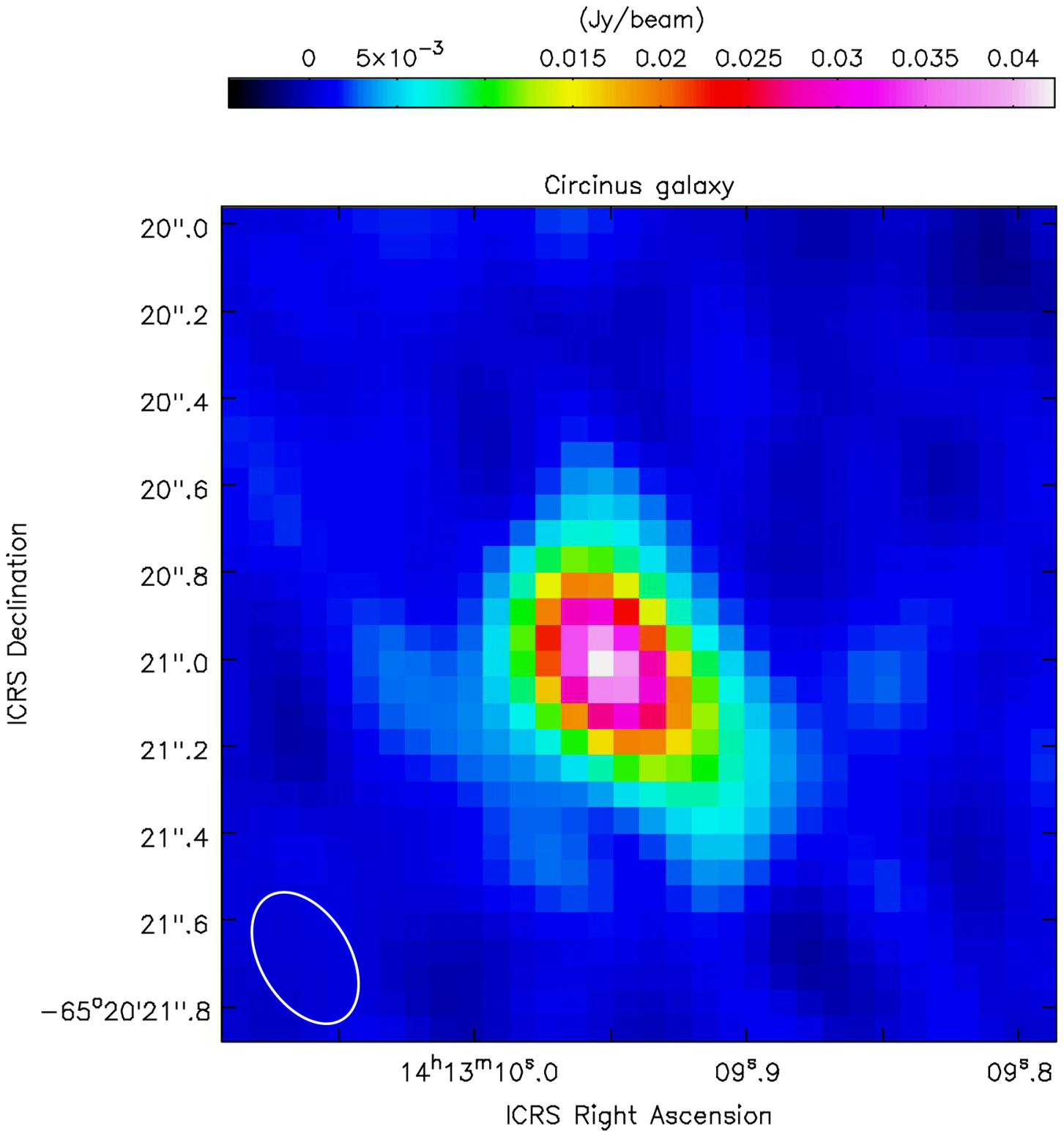}
\caption{Continuum maps of NGC\,4945 ($left$) and the Circinus galaxy
  ($right$) at 321\,GHz (Band~7).  The synthesized beams are shown in
  the bottom left in each panel (NGC\,4945: 0.381$\arcsec$$\times$0.313$\arcsec$;
  Circinus galaxy: 0.330$\arcsec$$\times$0.206$\arcsec$).  For both maps the flux
  densities are shown by color-scales in Jy\,beam$^{-1}$. \label{cont}}
\end{figure*}

Sub-millimeter 321\,GHz (Band\,7) continuum images of NGC\,4945 and
the Circinus galaxy, obtained by integrating over the full frequency range
(after the removal of the maser line emission), are shown in
Figure~\ref{cont}. The continuum map of NGC\,4945 shows resolved radio
structures, a bright radio source and extended components, which are
consistent with the map at the same frequency at 0.54$\arcsec$
angular resolution \citep{hagi16,pes16}, with a peak position of
$\alpha$(2000):13$^h$05$^{\rm m}$27.482, $\delta$(2000):
$-$49$\degr 28\arcmin 05\arcsec.40$.
The continuum map of the Circinus
galaxy shows a bright radio source in the center, accompanying 
extended structure tracing the spiral arms \citep{izu18}, with a peak
position of $\alpha$(2000):14$^h$13$^{\rm m}$09.954,
$\delta$(2000):$-$65$\degr 20\arcmin 21\arcsec.01$, which is also
consistent with that measured at lower angular resolution (0.54$\arcsec$)
within uncertainties of $\sim$0.01$\arcsec$ \citep{hagi13}.
The flux densities of the 321\,GHz continuum and the 321\,GHz \ho maser emission
are summarized in Table~\ref{tab3}.

All the maser features detected in our observations remain unresolved
at the angular resolution of $\sim$0.33$\arcsec$, or 6.9\,pc. The
position of the prominent blue-shifted feature peaking at
\vlsr\,=\,269.2\,\kms is
$\alpha$(J2000): 14$^{\rm h}$13$^{\rm m}$09$\fs$952,
$\delta$(J2000): $-$65$\degr$20$\arcmin$21.013$\arcsec$,
and that of the strongest feature at \vlsr\,=\,600.92\,\kms is
$\alpha$(J2000): 14$^{\rm h}$13$^{\rm m}$09$\fs$949,
$\delta$(J2000): $-$65$\degr$20$\arcmin$21.019$\arcsec$.
From the ALMA Cycle~4 Capabilities, the image registration accuracy is
$\sim${$\theta$}/{20.0}, where $\theta$ is the angular
resolution. Consequently, the registration accuracy is $\sim$0.33$\arcsec$/20
$\sim$ 0.0165$\arcsec$ in our observations and these positions of the bulk of
Doppler-shifted features between \vlsr\,=\,200 and 800\,\kms in
Figure~\ref{highv} are consistent within that accuracy.

\begin{figure*}[ht!]
 %%         \gridline{\fig{f5a.pdf}{0.3\textwidth}{(a)}
 %         \fig{f5b.pdf}{0.3\textwidth}{(b)}
 %       \fig{f5c.pdf}{0.3\textwidth}{(c)}
 %         }
%
%\includegraphics[angle=0,scale=0.3]{f5c.eps}  
%
\epsscale{1.25}
\plotone{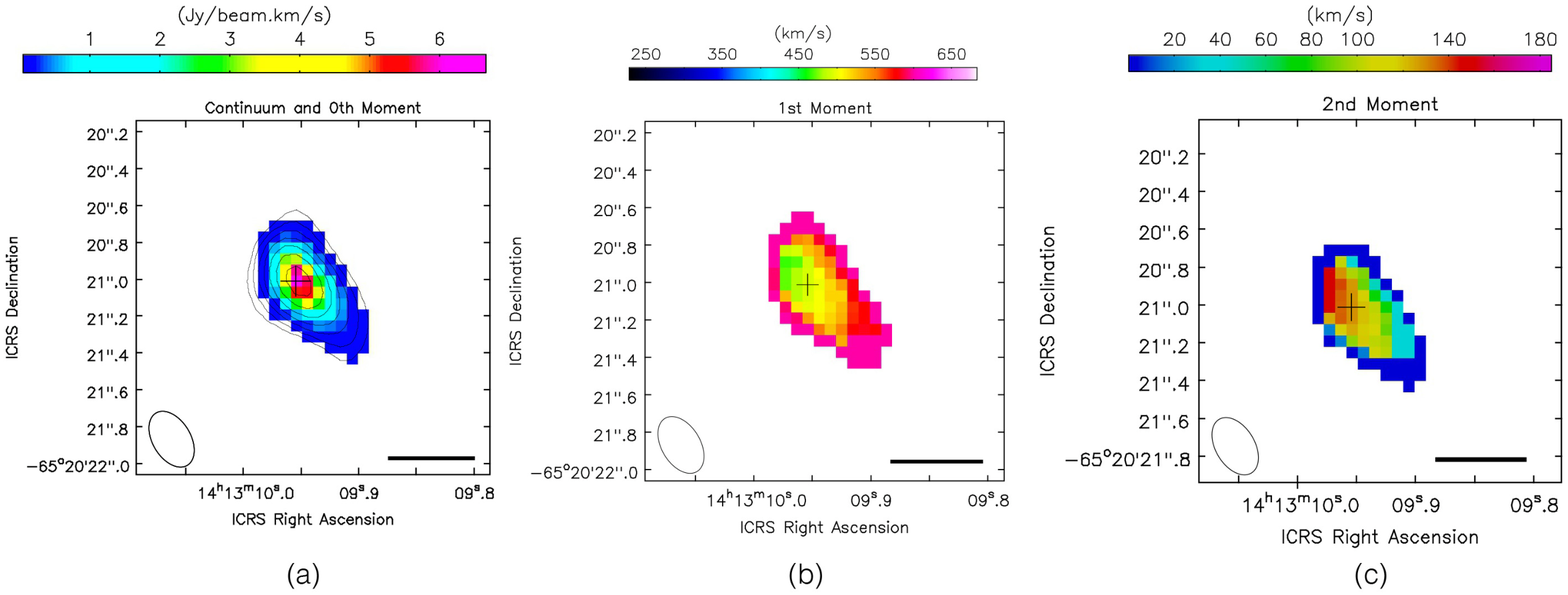}
\caption{Color-scale maps of 321\,GHz \ho maser emission in the Circinus galaxy:
  (a) Integrated line intensity map (0th moment) and 321\,GHz
  continuum with overlaying contours having a peak flux density of
  42.5\,\mb with the contour levels at 7, 9.8, 14, 19.7, 28, 39.5, and
  56\,$\sigma$ (1\,$\sigma$\,=\,0.63\,\mbbb), (b) mean velocity (first
  moment) map, and (c) velocity dispersion (second moment) map.
  These figures
  are obtained from our ALMA Cycle~4 observations. The total velocity
  coverage of the figures is \vlsr\,=\,228$-$685\,\kmss. The synthesized
  beam sizes (0.33$\arcsec$$\times$0.21$\arcsec$) are shown in the bottom left of each
  panel.  A cross in each panel marks the nucleus (the continuum peak)
  of the galaxy.  The horizontal bar at the bottom marks 10\,pc. Note that
  in (b) the surrounding components, reaching an extreme of $\sim$650\,\kms (pink)
  are \ho emission extending larger than the synthesized beam size, resulting in a ring-like appearance.
  The blueshifted maser features peaking at \vlsr\,$\simeq$\,269\,\kms in Figure~\ref{highv} ($left$) are also
  extended and do not show any velocity gradient on the map. \label{moment}}
\end{figure*}

Figure~\ref{moment} displays the velocity-integrated line intensity
(0th moment), overlaid on the 321\,GHz continuum (contour map), mean
velocity (first moment), and velocity dispersion (second moment) maps
for the 321\,GHz \ho maser in the Circinus galaxy. From the 0th moment map,
the source size of the maser emission deconvolved by the synthesized
beam is calculated to be 266$\pm$7\,mas $\times$ 158$\pm$2\,mas,
corresponding to 5.6\,pc\,$\times$\,3.3\,pc, at a position angle (PA) of
35$^{\circ}$ $\pm$ 1$^{\circ}$.
\setcounter{table}{0}

\begin{deluxetable*}{lccccccccc}
%\tablenum{2}
\tablecaption{Summary of the ALMA observations \label{tab1}}
\tablewidth{0pt}
\tablehead{
\colhead{Galaxy}&\colhead{RA}&\colhead{DEC}&\colhead{D}&\colhead{Date}&$N_{A}$&\colhead{$\theta_{b}$,PA}& \colhead{t$_{\rm on}$}&\colhead{$\sigma_{\rm c}$}&\colhead{$\sigma_{\rm l}$}\\
&\colhead{(J2000)}&\colhead{(J2000)}&(Mpc)&&   &\colhead{(arcsec$^2$, $^{\circ}$)}   &(min)&&
}
\decimalcolnumbers
\startdata
NGC 4945 &13$^{\rm h}$05$^{\rm m}$27$^{\rm s}$.279 & $-$49$^{\circ}$28${\arcmin}$04${\arcsec}$.44  &3.7&2016-11-18 &45&0.381$\times$0.313, -56.5&33&0.6&12.6 \\
\hline
Circinus   & 14$^{\rm h}$13$^{\rm m}$09$^{\rm s}$.906 & $-$65$^{\circ}$20${\arcmin}$20${\arcsec}$.46&4.2 &2017-05-06&47 &0.330$\times$0.206,      31.0 & 28&0.6&13.9\\
\enddata
\tablecomments{(2)-(3) Interferometry phase center positions in R.A.\ and Declination; (4) Adopted distances for NGC\,4945 from \citet{tul13} and for Circinus from \citet{fre77}; (5) Date of observations in YYYY-mm-dd format; (6) Number of 12-m antennas used in the ALMA observations; (7) Synthesized beam size and the beam position angle; (8) On-source time for the target source; (9)  Rms noise values in \mb in the clean continuum images; (10) Rms noise levels in \mb in a 488.3\,kHz (0.458\,\kmss) spectral channel in the clean line images.}
%\tablenotetext{a}{~Interferometry phase center positions in R.A.\ and Declination}
%\tablenotetext{b}{~Luminosity distances, adopted from NED}
%\tablenotetext{c}{~Date of observations in yyyy-mm-dd }
%\tablenotetext{d}{~Number of the 12-m antenna used in the ALMA observations}
%\tablenotetext{e}{~On-source time for the target source}
%\tablenotetext{f}{~Synthesized beam size}
%%\tablenotetext{g}{~The beam position angle}
%\tablenotetext{g}{~rms noise values in a 488.3 kHz spectral resolution in the clean image, depending on channel}
%
\end{deluxetable*}
%
%Tid70m Table2 revised
%
\begin{deluxetable*}{llccccc}
%\tablenum{2}
\tablecaption{Summary of the Tidbinbilla Observations \label{tab2}}
\tablewidth{0pt}
\tablehead{
\colhead{Galaxy}&\colhead{Date of Observation}&\colhead{$\Delta$V}&\colhead{$\Delta\nu$}&\colhead{T$_{\rm sys}$}&\colhead{t$_{\rm on}$}&\colhead{$\sigma_{\rm l}$} \\
&&(\kms)&(\kms)&(K)&(min)&(mJy)
}
\decimalcolnumbers
\startdata
NGC 4945 &2014-04-07$^{a}$, 2016-04-13, 2016-10-24&-250--1450&0.42&45--90&6--28&50--300 \\
 &2016-11-26, 2016-12-10&&& \\
\hline
Circinus Galaxy &2012-06-26, 2012-09-07$^{b}$, 2016-10-24&0--800&0.42&50--65&2--11&40--90\\
 &2017-05-06&&& \\
\enddata
\tablecomments{(2) Date of observations in YYYY-mm-dd format { ($a$: From published data in \citet{hagi16}, $b$:  from  \citet{hagi13})}; 
(3) Velocity range of 22\,GHz maser spectra in Figures $\ref{fign4945},\ref{circinus22}$; (4) Velocity resolution in each spectrum;  (5) Typical values of system noise temperature; {(6) Approximate on-source time for the target source}; (7) Rms noise levels in the 0.42\,\kms spectral channel in each spectrum. Note that the rms values have an uncertainty due to less accurate flux density estimation (see Section \ref{obs2}). The ATNF Correlator was used for the observations up to 2014 April 7, and the Smithsonian Astrophysical Observatory (SAO)
spectrometer was used for the subsequent observations.}
\end{deluxetable*}
\begin{deluxetable*}{rcccccccccc}
%\tablenum{2}
\tablecaption{Measured properties of 321\,GHz \ho Maser and Continuum  Emission toward Circinus galaxy and NGC\,4945 \label{tab3}}
\tablewidth{0pt}
\tablehead{
\colhead{Galaxy} &\colhead{Epoch$^{a}$}&\colhead{Date$^{b}$}& \multicolumn{3}{c}{Maser Line} & &\multicolumn{2}{c}{Continuum$^{d}$} \\
\cline{4-6}\cline{8-9}
&&& \colhead{Total Flux}&\colhead{Peak Flux}&\colhead{Peak Velocity$^{c}$} &&\colhead{Total Flux}&\colhead{Peak Flux}  \\
%&\colhead{(Line)}&\colhead{(Line)}&\colhead{(Line)} &\colhead{(Continuum)}&\colhead{ (Continuum)} &\\
&&&\colhead{(Jy)}&\colhead{(\jb)}&(\kmss)&&\colhead{(mJy)} &\colhead{(\mb)} &   
}
%\decimalcolnumbers
\startdata
Circinus &1&2012-06-03&0.35$\pm$0.04&0.13$\pm$0.01 &531.1 \kms &&55$\pm$6&40.6$\pm$4.1 \\
galaxy&2&2017-05-06&4.3$\pm$0.4&1.56$\pm$0.16& 601.6 \kms &&90.4$\pm$9.0&42.5$\pm$4.3 \\
\hline
NGC\,4945 &1&2012-06-03&0.052$\pm$0.005&0.040$\pm$0.004& 722.6 \kms &&543$\pm$54&136.5$\pm$13.7 \\
&2&2016-11-18&-&- &-&&475$\pm$48&61.6$\pm$6.2 \\
\enddata
\tablecomments{$a$: Epoch number;  $b$: Date of observations in YYYY-mm-dd format; $c$: Center velocities of a maser feature in optical LSR definition; $d$: 321\,GHz continuum emission from the nuclei of the galaxies}
\end{deluxetable*}
%% The "ht!" tells LaTeX to put the figure "here" first, at the "top" next
%% and to override the normal way of calculating a float position
 %%Spectra of \ho maser emission toward the nucleus of the Circinusgalaxy.
%  Vertical arrows indicate the LSR systemic velocity of
%  the Circinus galaxy, 433\,\kmss.
 %
%The 1\,$\sigma$ noise levels are $\sim$25--300\,mJy in a  0.42\,\kms velocity channel.
% obtained toward the region marked by a box on the 320\,GHz continuum image   47mJy * 1.5= 70 mJy
%
%restoringbeam': {'major': {'unit': 'arcsec', 'value': 0.33141028881073},
% 'minor': {'unit': 'arcsec', 'value': 0.20649318397045135},
%  'positionangle': {'unit': 'deg', 'value': 31.110733032226562
%

\section{DISCUSSION}\label{discussion}
Given the observed flux density variability of the maser features in the
spectra, it is clear that the 321\,GHz \ho emission in
NGC\,4945 and the Circinus galaxy occurs through the maser amplification
process.
%, although the maser in NGC\,4945 was not found in the second epoch.
Here we briefly discuss our non-detection of the 321\,GHz \ho
maser in NGC\,4945 and results of the image analysis of the 321\,GHz maser
in the Circinus galaxy.

\subsection{No Detection and Intensity Variability of  321\,GHz \ho Maser in NGC 4945}\label{sub:nomaser4945}
{From the two-epoch observations at 321\,GHz in 2012 and
2016 and five-epoch single-dish observations at 22\,GHz 
from 2014--2016, we find that strengths of the 321\,GHz and 22\,GHz
\ho masers vary in an apparently correlated fashion (Figure~\ref{fign4945}).
%although we have only two-epochs data of the 321\,GHz \ho maser.
The total and peak fluxes of the 321\,GHz continuum in 2016 have
decreased from those in 2012 (Table~\ref{tab3}). The detected
321\,GHz continuum emission of the galaxy is dominated by dust
emission and contributions from the synchrotron and free-free
emissions to the observed flux density are small as they decrease
rapidly at sub-millimeter wavelengths. The most dominant nuclear
continuum component in the center of the galaxy on the map traces
the dust lanes heated by the AGN in the obscured nucleus.  
% lying over the galactic disk 

%The combination of our two-epoch observations of the 321\,GHz maser
%emission and the five-epoch observations of the 22\,GHz maser emission
%shows
 {Although we have only two epochs of data,}
there is an apparent correlation between the strength of the masers and
the 321\,GHz radio continuum emission from the obscured nucleus
(Table~\ref{tab3}).
This is consistent with the non-detection of 
321\,GHz maser emission in 2016. The 321\,GHz maser emission that was
detected in 2012 {and higher intensity of the 22\, GHz maser emission in 2014} were most likely to be enhanced by increased
AGN-activity through the thick dust lane in our line of sight around
that period. This supports the conclusion that the 321\,GHz \ho emission in
the galaxy is a nuclear maser.

{If the maser is powered by X-ray emission from an AGN \citep[e.g.,][]{neu94},
then the observed intensity variability of the 22 and 321\,GHz maser
could be due to the central engine \citep[e.g.,][]{mal02}.
In the galaxy, there is a luminous nuclear hard X-ray source (X-ray luminosity: $L_X \sim 4 \times 10^{43}$\,erg s$^{-1}$)
buried in the Compton-thick nucleus \citep[e.g.,][]{mar17}. The X-ray luminosity is highly variable with a time scale of $\sim$1\,day,
suggesting that the size scale of X-ray emitting region  
is no more than $\sim$0.001\,pc \citep{mad00}, however it is uncertain whether this compact X-ray region is responsible for the
decreased maser intensities over the four years. Moreover, little is known about the variability of the 321\,GHz masers.}
\subsection{The 321\,GHz Maser in Circinus galaxy}\label{sub:maserCG}
\subsubsection{Origin of the  maser and physical conditions giving rise to the maser}
The isotropic luminosity of the 321\,GHz maser emission, assuming
isotropic radiation of the maser, is calculated to be
$\sim$65 \lsun using the formula, \\
$L$\,=\,1.04$\times$10$^{-3}$\,$\nu$\,$D^2$\,$\int$$S\,dv$\,$\lsun$,
where $\nu$ is the rest frequency (321.226\,GHz),
$D$ is the distance to the Circinus galaxy of\,4.2 Mpc \citep{fre77}, and
$\int S \,dv$ is the integrated intensity (11.1\,Jy \kmss) \citep{liz16}.
This value is roughly a factor of two larger than the
isotropic luminosity of the 22\,GHz maser emission of $\sim$32\,$\lsun$ in
the galaxy \citep{whi86} (adopting $D$\,=\,4.2\,Mpc).
This strongly supports the belief that the maser in the galaxy is a nuclear
  maser arising from AGN-activity and not from star-forming activity
  in the galaxy.  Moreover, the position of the 321\,GHz maser
coincides with the radio nucleus and the significant flux variability
is measured between the two epochs, which indicates that the maser 
may be associated with ejecta from the active nucleus such as a radio jet.
Given the relatively higher maser luminosity, comparable with
that of the 22\,GHz megamaser known in the galaxy, we conclude that
the maser amplifies the background continuum and its variability is
due to the activity from a central engine in the active nucleus
and so the observed 321\,GHz maser in the Circinus can be
categorized as a luminous \ho megamaser.  

{The isotropic luminosity of the 183\,GHz \ho maser in NGC\,4945,
$\sim$1300\,$\lsun$ is roughly two orders of magnitude larger than
that of the 321\,GHz maser in NGC\,4945 at $\approx$5 \lsun~and
the Circinus galaxy at $\approx$ 65--100 $\lsun$ \citep[this paper;][]{liz16, hagi16,
  pes16}.} The upper state energies ($E_u/k$) of the \ho emission for the 22\,GHz and
183\,GHz transition are 644\,K and 205\,K, respectively, while $E_u/k$
for the 321\,GHz transition is much higher: 1862\,K.
In NGC\.4945, the energy of a significant portion
of dense gas giving rise to maser amplification may not be
sufficiently high enough for intense 321\,GHz maser to arise, while it
could be sufficient for elevating the strong 183\,GHz maser
emission. Therefore, insufficient excitation of the 321\,GHz maser in
NGC\,4945 could make the strength of the maser weaker than that of the
183 and 22\,GHz masers \citep{liz16}. 
 It is therefore most likely that the strength of the 183\,GHz maser in the Circinus galaxy will be
  greater than the 321\,GHz maser in the galaxy in future observations.
However, due to the
higher excitation condition required for the 321\,GHz maser, it is more
likely that the 321\,GHz masers
%in the Circicus galaxy and NGC\,4945
can probe regions closer to the nucleus, which can be verified
with higher angular resolution observations.
%It should be noted that the strengths of both the 321\,GHz continuum and \ho maser emission increased from 2011 to 2017 (table~\ref{tab2}), suggesting that the maser excitation relates to the activity of an active galactic nucleus as well as the case 
%in NGC\,4945.
%The most Doppler-shifted velocity of 460 \kms w.r.t the systemic velocity \citep{linc03a}, while those of the 321\,GHz transition from $\sim$ 270 \kms to $\sim$ 670 \kms in our observation, excepting the minor detection. This 
%with a radius of ??\,pc
%0.3x5x10^-6x4.2 10^6=1.5x4.2=6.3 , 0.2x5x10^-6x4.2 10^6=1.5x4.2=4.2
%#ant C0:18   18*17/2=153
%##ant C4 :47    47*46/2=1081     1081/153 = 7.06

\subsubsection{Velocity Gradient}\label{sub:gradient}
One of the most exciting results from our observations is that part of
the velocity field of the dense molecular gas within $\sim$6\,pc
of the nucleus in the Circinus galaxy has been resolved at our $\sim$0.3$\arcsec$
spatial resolution. As is evident in Figure~\ref{moment}(b), there is a distinct
velocity gradient in the 321\,GHz \ho maser emission spanning $\sim$110\,\kms
in the first moment map.
Figure~\ref{moment}(c) shows the second moment map which, as for Figure~\ref{moment}(b),
is produced from the spectral-line cube containing maser features lying
at \vlsr\,=\,256--670\,\kmss. This indicates that the maser
emission is partly resolved in our second epoch observation at
$\sim$0.3$\arcsec$ ($\sim$6\,pc) resolution, 
although neither
velocity gradient nor velocity dispersion was found in the first epoch
observation in 2011 \citep{hagi13}. This is likely to have arisen from the
improvement in $(u, v)$ coverage due to the significantly larger number of
baselines, as a result of the number of antennas increasing from 18 to 47,
and the longer integration time of 28 minutes, compared to 18 minutes in the first epoch. 

Assuming that the observed masers are on a simple circular disk
with a rotational velocity of V$_{rot}$ at a radius $r$ from the
nucleus, the observed velocity gradient ($dV$/$dl$) along PA\,=\,35$\degr$
can be expressed as
$dV$/$dl$ $\simeq$$d$(V$_{rot}$~sin$\theta$~sin$i$)/$d$($r$$\theta$) $\simeq$ $V_{rot}$~sin$i$/$r$,
where $l$(=$r$$\theta$) is the projected distance,
$i$ is the disk inclination, and the approximation of
sin$\theta \simeq \theta$ is being made.
Adopting the disk inclination of
$>$75$\degr$ (sin$i$ $\simeq$1) estimated from the near-infrared
emission \citep{tri14} or the CO disk with $r$\,=\,100--600\,pc
\citep{cur08}, together with the observed parameters of
$dV/dl$ = 8.8\,\kms\,pc$^{-1}$ and
$V_{red} - V_{sys}$ = 682\,\kms--433\,\kms = $V_{rot}$ $\simeq$ 250\,\kmss,
$r$ is estimated to be $\sim$28\,pc.
{Since the most red- or blue-shifted features are interpreted as
originating in an edge-on disk \citep[e.g.,][]{miyo95}, the
rotation speed ranges up to $\sim$250\,\kms using the most
Doppler-shifted velocity of \vlsr $\simeq$682\,\kmss.  This value is
to the ``lower'' rotation velocity ($\sim$260\,\kmss) of the
sub-parsec scale disk in the Circinus galaxy, estimated from the 22\,GHz
maser emission \citep{mac09}, compared to the Keplerian rotation velocity
($\sim$1100\,\kmss) of the disk in NGC\,4258.}  Considering a simple
circular motion around the nucleus, we estimate the dynamical mass
(M$_{\rm{dyn}}$) confined within the central 28\,pc to be
$\sim$4.3$\times$10$^8$\,\msun, which is approximately two orders of magnitude
larger than the enclosed mass of $\sim$1.7$\times$10$^6$\,\msun
within $r$\,=\,0.1\,pc that was measured by VLBI observations, assuming a
circular motion \citep{linc03b}. 

We note that the derived value of the dynamical mass (r\,$\la$\,28\,pc) is
comparable with that of the larger scale dynamical mass of
$\sim$3.2$\times$10$^8$\,\msun within the central 140\,pc \citep{cur98} and
approximately two orders of magnitude larger than the total molecular
mass of M$_{\rm{H_2}} \approx$\,3$\times$10$^6$\,\msun  within the
central $\sim$75\,pc in the circumnuclear disk probed by CO (3--2) molecular gas
with the deconvolved source size of 3.62$\arcsec$\,$\times$\,1.66$\arcsec$
($\sim$76\,pc\,$\times$\,35\,pc) \citep{izu18}. This gives a gas mass
fraction of only M$_{\rm{H_2}}$/M$_{dyn}$\,$\approx$\,0.6\% within the
central 28\,pc.

For the Circinus galaxy, a molecular gas fraction in the central 500\,pc
of M$_{\rm{H_2}}$/M$_{\rm{dyn}}$ $\approx$ 50\% 
has been reported \citet{cur08}. It is suggested that the fraction of the gas
mass is up to 25\% of the dynamical mass in spiral galaxies
\citep{you82}.

%This estimation suggests that {\bf the observed 321\,GHz \ho masers does not
%probe a disk on a scale of 10\,pc.}

{This discrepancy is probably caused by our assumption that the 321\,GHz
maser traces the outer part of the central disk within the central
$\sim$28\,pc, which would result in an overestimate of the dynamical mass on a
scale of 10\,pc. 

If the 321\,GHz maser resides in the 22\,GHz maser disk with $r\,\simeq$\,0.1--0.4\,pc, obtained by
VLBI mapping of the 22\,GHz maser \citep{linc03b, mac09},  the 321\,GHz masers should be located
closer to the central engine  than the 22\,GHz masers as the 321\,GHz maser requires higher temperatures than 22\,GHz masers.
For the 321\,GHz masers, strong inversion of the emission occurs in denser regions at a gas kinetic temperature, T$_k$ $\ga$ 1000\,K 
and n(H$_2$) = 10$^8$--10$^{10}$ cm$^{-3}$. For the 22\,GHz masers,  the strong inversion occurs over a broader range of physical conditions, 
spanning  T$_k$ of $\sim$300--1000 K and  n(H$_2$) $\la$10$^8$ to 10$^{10}$\,cm$^{-3}$ \citep{eli89,yat97}.}

We thus postulate that the 321\,GHz maser traces the inner part of the masering disk with  $r$\,=\,0.1--0.4\,pc, as in the circumnuclear disk illuminated by X-rays the temperature increases closer to the central engine \citep[e.g.,][]{neu95}.
Alternatively, the observed \ho velocity gradient could trace non-circular motions and may not be connected to the overall velocity field on that scale.

Since the position angle of the
detected gradient is comparable to that of the synthesized beam (see
Table~\ref{tab3}), the intrinsic position angle of the velocity
gradient is unknown. However, the position angle
(35$\degr$$\pm$1$^{\circ}$) of the maser emission deconvolved by the
synthesized beam largely agrees with that of the 22\,GHz \ho maser disk
(29$^{\circ}$$\pm$3$^{\circ}$ at $r$\,=\,0.1\,pc, and
56$^{\circ}$$\pm$6$^{\circ}$ at $r$\,=\,0.4\,pc \citep{linc03b}.
This suggests that the inner part of each gas structure has 
similar position angles on all spatial scales \citep{cur08}.

Further observations at higher angular
resolution would enable us to clarify the inner part of gaseous
structure and the true origin of the velocity gradient.
\subsubsection{22\,GHz Maser and 321\,GHz Maser Comparison}\label{sub:masercomparison}
The VLBI study of the 22\,GHz maser emission by \citet{linc03b}
revealed the spatial distribution of the individual maser features
within the total extent of $\sim$0.04$\arcsec$\,$\times$\,0.09$\arcsec$
($\sim$0.8\,$\times$\,1.9\,pc) and the origins of those maser sources were divided
into two categories: the thin disk or
outflows in the galaxy. They found that the velocity range of the
maser emission associated with outflows is up to $\sim$\,$\pm$160\,\kms
from the systemic velocity, which corresponds to \vlsr~$\simeq$
300--570\,\kmss, while the disk maser emission spans 
\vlsr~$\simeq$ 250--280 \kms in the blueshifted velocity range and
\vlsr~$\simeq$ 570--700 \kms in the redshifted range.
The bulk of the emission in the 321\,GHz maser
spectra has been detected in blue- and red-ward velocities but not at
\vlsr~$\simeq$ 300--500 \kmss, which  implies that the 321\,GHz maser
features in the blue- and redshifted velocity ranges largely
originate from the thin disk, as in the case of the 22\,GHz masers
revealed by the VLBI study. However, in the case of the 321 GHz maser,
it is not possible to distinguish 
whether the maser originates from a disk, an outflow, or
both a disk and outflow due to lack of spatial resolution of our ALMA observations.

Contrary to the disk maser, on 2017 May 6
(Figure~\ref{circinus321} and Figure~\ref{highv}(left)), strong features centered on \vlsr\,=\,359.29\,\kms
appeared in the 321\,GHz spectrum, while blueshifted
features centered around \vlsr\,=\,380\,\kms were found in the 22\,GHz
maser spectrum.  We interpret this to mean that these blueshifted features, that
appeared at both 321\,GHz and 22\,GHz, are outflow masers that flared at
the same period and they are associated with significant shocks with
the surrounding molecular materials in the nuclear region. This also
supports the hypothesis that the 321\,GHz and 22\,GHz masers are powered by the central engine, as argued in Section~\ref{sub:nomaser4945}. 

It should be noted that submillimeter stellar \ho masers are often
found to be associated with outflow jets, while the 22\,GHz masers are primarily found 
in disks \citep[e.g.,][]{pat07}. However,  it still remains unclear to what extent
these findings can be applied to the cases of extragalactic 321 GHz \ho masers.
%It is interesting note that the direction and are  to HI kpcs scale gradient. 
%0.04 x 5 x 10^-6 x 4.2 x 10^6 =0.2 x 4.2=0.84
%0.09 x  5 x 10^-6 x 4.2 x 10^6 = 0.45 x 4.2=1.89
%We cannot see warping in the gradient map. \\high-dispersion region at offset ?0 3 north of the S nucleus.

%One may wonder where the maser arises from?
\subsubsection{Velocity Dispersion}\label{sub:dispersion}
%462-433=29, V/sigma = 29/148=0.1959,    575-433/98=142/98
In Figure~\ref{moment}(c), we find a high velocity dispersion
region, with a velocity dispersion ($\sigma_{v}$) of $\sim$148\,\kmss,
{where the rotation velocity (V$_{\rm{rot}}$ = \vlsr --
\vsys) is 29\,\kms in the first moment map, is offset to the northeast of
the continuum nucleus and a gradient of the velocity dispersion is
seen between \vlsr $\simeq$ 98\,\kms and 148\,\kms along the southwest to
northeast direction.  }
We speculate that velocities at the
northeastern part are more blueshifted than those at the southeast
part in the first moment map, which reduces the opacity and results in
making the turbulent molecular gas appear stronger
\citep{ima18,ima20}.
It is interesting to note that the relative
velocity dispersion values of V$_{\rm{rot}}$/$\sigma_{v}$
\citep{tre20} increase from $\sim$ 0.2 to 1.4 from the northeast to
the southwest direction and the velocity dispersion distribution is
similar to the velocity gradient. It is not certain whether the
velocity dispersion distribution can be explained by our simple
speculation, however it could be dynamically linked to
%organized movements such as
circular or non-circular motion on scales of 10\,pc.

Future sub-millimeter VLBI mapping of the 321\,GHz \ho masers
at milliarcsecond resolution \citep[e.g.,][]{fis13} should be able to
clarify the sub--parsec-scale kinematics of the galaxy by pin-pointing
the distributions of each of the maser features.
\section{Conclusions}\label{summary}
We have observed \ho masers in the 321\,GHz transition
toward the nearby active galactic nuclei NGC\,4945 and the Circinus galaxy
at $\sim$0.3$\arcsec$ angular resolution with ALMA Band~7.
Comparison of two epochs
of ALMA observations show that the flux densities of the 321\,GHz \ho
masers in the galaxies are significantly variable on timescales of 
4--5 years and the nuclear continuum emission varies in a correlated
fashion with the maser in NGC\,4945, while the correlation between the
maser and the nuclear continuum is less certain in the Circinus galaxy.
The significant variability in the flux densities has confirmed that
the observed \ho emission arises from nuclear masers.  With the 70\,m telescope
at Tidbinbilla we also conducted single-dish measurements of the 22\,GHz
maser emission over 3--5 epochs between the two ALMA observations
toward the galaxies to compare the variability of the fluxes and
velocities between the 321\,GHz and 22\,GHz maser lines.
This revealed in the Circinus galaxy the flaring features of the 321\,GHz and 22\,GHz
maser emission were found largely in the same blueshifted velocity range on
the same day. It is thus likely that the 321\,GHz and 22\,GHz masers 
originate in similar kinematics and physical conditions.

Contrary to the previous lower spatial resolution observations,
made with a much smaller number of antennas, in the Circinus galaxy we
detected a velocity gradient and dispersion, traced by the 321\,GHz
\ho maser emission, in a roughly southwest to northeast direction.
This position angle is similar to that of the sub-parsec scale thin disk
probed by the 22\,GHz \ho masers. With the velocity
gradient spanning $\sim$110\,\kmss, we estimate the dynamical mass
within the central 28\,pc to be $\sim$4.3$\times$10$^8$\,\msun,
which is significantly larger than the dynamical mass and
the molecular gas mass within the similar radii.
Therefore,  the observed velocity gradient does not likely trace 
the extension of the sub-parsec--scale 22\,GHz maser disk but might indicate
non-circular motion rather than the simple circular motion of the disk on
that scale. If the 321\,GHz maser is in the circumnuclear disk,
it should be observed closer to the central engine than the 22\,GHz masers as 
the excitation of 321\,GHz maser needs a higher temperature in denser gas.

We speculate that blueshifted \ho maser emission at north-eastern
region reduces the opacity, making turbulence of molecular gas more
conspicuous, which results in the higher velocity
dispersion at that region. The gradient of the velocity dispersion is
also found along the same position angle as the velocity gradient. Its
origin is uncertain from our data, however it could be physically
connected to organized movements like circular motion.
Overall, the comparison of our data to the earlier VLBI
studies of the 22\,GHz \ho maser suggests that the 321\,GHz \ho maser
  could not trace an extension of the sub-parsec scale thin disk but the
  variable blue-shifted features might trace 
the nuclear outflows.
Higher spatial observations using ALMA or future sub-millimeter
VLBI might reveal whether the 321\,GHz \ho
maser probes a dynamical structure of inner parsecs or inner part of the circumnuclear disk in the
\object{Circinus galaxy}.
\acknowledgments 
We thank the anonymous referee for useful comments. YH thanks Kazuya Saigo for his support with the ALMA data reduction pipeline. This work was supported by JSPS KAKENHI Grant Number
JP15H03644 (YH) and 21K03632 (MI).  This paper makes use of the following ALMA data:
ADS/JAO.ALMA \#2016.1.00150.S, \#2011.0.00121.S. ALMA is a partnership of ESO, NSF and
NINS, together with NRC, MOST and ASIAA, and KASI, in cooperation with
the Republic of Chile. The Joint ALMA Observatory is operated by ESO,
AUI/NRAO and NAOJ. The Tidbinbilla 70-m telescope, DSS-43,  is part of the NASA
Deep Space Network and is operated by CSIRO. This research has made
use of the NASA/IPAC Extragalactic Database (NED), which is funded by
the National Aeronautics and Space Administration and operated by the
California Institute of Technology. This project made use of the
Smithsonian Astrophysical Observatory 4$\times$32k-channel
spectrometer (SAO32k) and the TAMS observatory Ctrl observing system,
which were developed by L.\ Greenhill (Center for Astrophysics), I.\ Zaw
(New York University Abu Dhabi), D.\ Price, and D.\ Shaff, with funding
from SAO and the NYUAD Research Enhancement Fund and in-kind support
from the Xilinx University Program. 

%
%% Following the acknowledgments section, use the following syntax and the
%% \facility{} or \facilities{} macros to list the keywords of facilities used 
%% in the research for the paper.  Each keyword is check against the master 
%% list during copy editing.  Individual instruments can be provided in 
%% parentheses, after the keyword, but they are not verified.

%\vspace{5mm}
\facilities{ALMA, Tidbinbilla 70m (DSS-43)}

%% Similar to \facility{}, there is the optional \software command to allow 
%% authors a place to specify which programs were used during the creation of 
%% the manusscript. Authors should list each code and include either a
%% citation or url to the code inside ()s when available.

\software{CASA, AIPS, ASAP, GBTIDL}

%% The reference list follows the main body and any appendices.
%% Use LaTeX's thebibliography environment to mark up your reference list.
%% Note \begin{thebibliography} is followed by an empty set of
%% curly braces.  If you forget this, LaTeX will generate the error
%% "Perhaps a missing \item?".
%%
%\newpage

%% This command is needed to show the entire author+affilation list when
%% the collaboration and author truncation commands are used.  It has to
%% go at the end of the manuscript.
%\allauthors

%% Include this line if you are using the \added, \replaced, \deleted
%% commands to see a summary list of all changes at the end of the article.
%\listofchanges


\begin{thebibliography}{}
%
\bibitem[Ar{\'e}valo et al.(2014)]{are14}
Ar{\'e}valo, P., Bauer, F.~E., Puccetti, S., et al.\ 2014, \apj, 791, 81
%
\bibitem[Braatz et al.(1994)]{bra94}
Braatz, J.~A., Wilson, A.~S., \& Henkel, C.\ 1994, \apjl, 437, L99
% doi:10.1086/187692
\bibitem[Braatz et al.(1996)]{bra96}
Braatz J. A., Wilson A. S., \& Henkel C. 1996, \apjs, 106, 51
%
\bibitem[Braatz et al.(2003)]{bra03}
Braatz, J. A., Wilson, A. S., Henkel, C., Gough, R., \& Sinclair, M. 2003, \apjs, 146, 249
% nearby AGN 8 H2OMM
\bibitem[Braatz \& Gugliucci(2008)]{bra08}
Braatz, J.~A., \& Gugliucci, N.~E.\ 2008, \apj, 678, 96
%. doi:10.1086/529538
%Arp220
\bibitem[Cernicharo et al.(2006)]{cer06}
Cernicharo, J., Pardo, J. R., \& Weiss, A. 2006, \apj, 646, L49
%
\bibitem[Curran et al.(1998)]{cur98}
Curran, S. J., Johansson, L. E. B., Rydbeck, G.,  \&  Booth, R. S. 1998, \aap, 338, 863
%
\bibitem[Curran et al.(1999)]{cur99}
Curran, S. J., Rydbeck, G.,  Johansson, L. E. B., \&  Booth, R. S. 1999, \aap, 344, 767
%
\bibitem[Curran et al.(2008)]{cur08}
Curran, S. J., Koribalski, B. S., \& Bains, I. 2008, \mnras, 389, 63
%
\bibitem[de Vaucouleurs et al.(1991)]{dev91}
de Vaucouleurs, G., de Vaucouleurs, A., Corwin, H. G., et al.\
%Buta, R. J., Paturel, G., \& Fouque, P.
1991, Third Reference Catalog of Bright Galaxies (Berlin: Springer)
%
\bibitem[Dos Santos \& L\'{e}pine(1979)]{dos79}
Dos Santos, P. M., \& L\'{e}pine, J. R. D. 1979, \nat, 278, 34
%
% h2o in SF regions
\bibitem[Elitzur et al.(1989)]{eli89} Elitzur, M., Hollenbach, D.~J., \& McKee, C.~F.\ 1989, \apj, 346, 983
%
\bibitem[Fish et al.(2013)]{fis13}
Fish, V.,  Alef, W.,  Anderson, J., et al.\ 2013, arXiv:1309.3519
%
\bibitem[Freeman et al.(1977)]{fre77}
Freeman, K. C., Karlsson, B., Lynga, G., et al.\ 1977, \aap, 55, 445
%new HV gas in Circinus\
\bibitem[Galametz et al.(2016)]{gala16}
Galametz, M., Zhang, Z.-Y., Immer, K., et al.\ 2016, \mnras, 462, L36
%rapid variation of maser from circinus
\bibitem[Greenhill et al.(1997a)]{linc97a}
Greenhill, L. J., Ellingsen, S. P., Norris, R. P., et al.
1997a, \apj, 474, L103
%ngc4945 VLBA
\bibitem[Greenhill et al.(1997b)]{linc97b}
Greenhill, L. J.,  Moran, J. M., \& Herrnstein, J. R. 1997b, \apj, 481, L23
%maser survey 7 discoveries
\bibitem[Greenhill et al.(2003a)]{linc03a}
Greenhill, L.~J., Kondratko, P.~T., Lovell, J.~E.~J., et al.\ 2003a, \apjl, 582, L11
%doi:10.1086/367602
%VLBI maser map Circinus
\bibitem[Greenhill et al.(2003b)]{linc03b}
Greenhill, L. J., Booth, R.~S., Ellingsen, S.~P., et al. 2003b, \apj, 590, 162
%disc maser survey
\bibitem[Greenhill et al.(2009)]{linc09}
Greenhill, L.~J., Kondratko, P.~T., Moran, J.~M., et al.\ 2009, \apj, 707, 787
%doi:10.1088/0004-637X/707/1/787
%
\bibitem[Hagiwara et al.(2002)]{hagi02}
Hagiwara, Y., Diamond, P. J.,  \& Miyoshi, M. \ 2002, \aap, 383, 65
%N4051
\bibitem[Hagiwara et al.(2003)]{hagi03} Hagiwara, Y., Diamond, P.~J., Miyoshi, M., et al.\ 2003, \mnras, 344, L53
%doi:10.1046/j.1365-8711.2003.07005.x
%
\bibitem[Hagiwara et al.(2013)]{hagi13}
%
Hagiwara, Y., Miyoshi, M., Doi, A., \& Horiuchi, S.\ 2013, \apjl, 768, L38
%
\bibitem[Hagiwara et al.(2016)]{hagi16}
Hagiwara, Y.,  Horiuchi, S., Doi, A., Miyoshi, M., \& Edwards, P. G. \ 2016, \apj, 827, 69
%
\bibitem[Hagiwara et al.(2018)]{hagi18}
Hagiwara, Y., Doi, A., Hachisuka, K.,  \& Horiuchi, S. \ 2018, \pasj, 70, 54
%
\bibitem[Henkel et al.(2005)]{hen05} Henkel, C., Peck, A.~B., Tarchi, A., et al.\ 2005, \aap, 436, 75
%doi:10.1051/0004-6361:20042175
%
\bibitem[Humphreys et al.(2005)]{liz05}
Humphreys, E. M. L., Greenhill, L. J., Reid, M. J., et al.\
%Beuther, H., Moran, J. M., Gurwell, M., Wilner, D. J., Kondratko, P. T.
2005, \apj, 634, L133
%
\bibitem[Humphreys et al.(2016)]{liz16}
Humphreys, E. M. L.,  Vlemmings, W. H. T., Impellizzeri, C. M. V.,
%Galametz, M., Olberg, M., Conway,  J. E.,
et al.\  2016, \aap, 592, L13
%disktorus
\bibitem[Imanishi et al.(2018)]{ima18}
Imanishi, M., Nakanishi, K., Izumi, T., et al.\ 2018, \apjl, 853, L25
%
\bibitem[Imanishi et al.(2020)]{ima20}
Imanishi, M., Nguyen, D.~D., Wada, K., et al.\ 2020, \apj, 902, 99
%183maser
\bibitem[Imanishi et al.(2021)]{ima21}
Imanishi, M., Hagiwara, Y., Horiuchi, S., et al.\ 2021, \mnras, 502, L79
%
\bibitem[Isobe et al.(2008)]{iso08}
Isobe, N., Kubota, A., Makishima, K., et al.\
%Gandhi, P., Griffiths, R. E., Dewangan, G. C., Itoh, T., Mizuno, T.
2008, \pasj, 60, 241
\bibitem[Iwasawa et al.(1993)]{iwa93}
Iwasawa, K., Koyama, K., Awaki, H., et al. 1993, \apj, 409, 155
%
\bibitem[Izumi et al.(2018)]{izu18}
Izumi, T., Wada, K.,  Fukushige, R.,  Hamamura, S., \& Kohno, K.  2018, \apj, 867, 48
%
%
\bibitem[Kondratko et al.(2006)]{kon06}
Kondratko, P. T., Greenhill, L. J., Moran, J. M.,
%Lovell, J. E. J., Kuiper, T. B. H.
et al.\ 2006, \apj, 638, 100
%
\bibitem[K\"{o}nig et al.(2017)]{kon17}
K\"{o}nig, S., Mart\'{i}n, S., Muller, S.,
%Cernicharo, J.,  Sakamoto, K.,
et al.\ 2017, \aap, 602, A42
%
\bibitem[Kuiper et al.(2019)]{kui19}
Kuiper, T.~B.~H., Franco, M., Smith, S., et al.\ 2019, Journal of Astronomical Instrumentation, 8, 1950014
%
\bibitem[Madejski et al.(2000)]{mad00} Madejski, G., {\.Z}ycki, P., Done, C., et al.\ 2000, \apjl, 535, L87
%
%
\bibitem[Maloney(2002)]{mal02} Maloney, P.~R.\ 2002, \pasa, 19, 401
%
\bibitem[Marconi et al.(1994)]{mar94}
Marconi. A.,   Moorwood, A. F.  M., Origlia, L., \& Olivia, E.\ 1994, Messenger, 78, 20
%
%
\bibitem[Marinucci et al.(2017)]{mar17} Marinucci, A., Bianchi, S., Fabbiano, G., et al.\ 2017, \mnras, 470, 4039
%
\bibitem[McCallum et al.(2005)]{mac05}
McCallum, J. N., Ellingsen, S. P., Jauncey, D. L., Lovell, J. E. J., \& Greenhill, L. J.\ 2005, \aj, 129, 1231
%
\bibitem[McCallum et al.(2009)]{mac09}
McCallum, J. N., Ellingsen, S. P.,  Lovell, J. E. J., Phillips, C. J., \& Reynolds, J. E.\ 2009, \mnras, 392, 1339
%
%
\bibitem[McMullin et al.(2007)]{mcm07} McMullin, J.~P., Waters, B., Schiebel, D., et al.\ 2007, Astronomical Data Analysis Software and Systems XVI, 376, 127
%
\bibitem[Miyoshi et al.(1995)]{miyo95}
Miyoshi, M.,  Moran, J., Herrnstein, J., et al.\
%Greenhill, L., Nakai, N., Diamond, P., Inoue, M.
1995, \nat, 373, 127
%
\bibitem[Nakai et al.(1995)]{naka95}
Nakai, N., Inoue, M., Miyazawa, K., Miyoshi, M., \& Hall, P. 1995, PASJ, 47, 771
%
\bibitem[Neufeld et al.(1994)]{neu94} Neufeld, D.~A., Maloney, P.~R., \& Conger, S.\ 1994, \apjl, 436, L127
%
\bibitem[Neufeld \& Maloney(1995)]{neu95} Neufeld, D.~A. \& Maloney, P.~R.\ 1995, \apjl, 447, L17
%
\bibitem[Patel et al.(2007)]{pat07} Patel, N.~A., Curiel, S., Zhang, Q., et al.\ 2007, \apjl, 658, L55
%
\bibitem[Pesce et al.(2016)]{pes16}
Pesce, D. W.,  Braatz, J. A., \& Impellizzeri, C. M. V. \ 2016, \apj, 827, 68
%
\bibitem[Puccetti et al.(2014)]{puc14}
Puccetti, S., Comastri, A., Fiore, F., et al.\ 2014, \apj, 793, 26
%
%
%
\bibitem[Tarchi et al.(2011)]{tar11}
Tarchi, A., Castangia, P., Columbano, A., et al.\ 2011, \aap, 532, A125
%. doi:10.1051/0004-6361/201117213
\bibitem[Treister et al.(2020)]{tre20}
Treister, E., Messias, H., Privon, G.~C., et al.\ 2020, \apj, 890, 149. doi:10.3847/1538-4357/ab6b28
%
\bibitem[Tristram et al.(2014)]{tri14}
Tristram, K. R. W., Burtscher, L., Jaffe, W., et al.\ 2014, \aap, 563, A82
%
%
\bibitem[Tully et al.(2013)]{tul13}
Tully, R. B., Courtois, H. M., Dolphin, A. E. et al.\ 2013, \aj, 146, 86
%
\bibitem[van Moorsel, Kemball, \& Greisen(1996)]{van96} van Moorsel, G., Kemball, A., \& Greisen, E.\ 1996, Astronomical Data Analysis Software and Systems V, 101, 37
\bibitem[Whiteoak et al.(1986)]{whi86} 
Whiteoak, J. B., \& Gardner, F. F. 1986, \mnras, 222, 513
%circinus conp
\bibitem[Yang et al.(2009)]{yan09}
Yang, Y., Wilson, A.~S., Matt, G., et al.\ 2009, \apj, 691, 131
%
\bibitem[Yates et al.(1997)]{yat97} Yates, J.~A., Field, D., \& Gray, M.~D.\ 1997, \mnras, 285, 303
%
\bibitem[Young and Scoville(1982)]{you82}
Young, J. S., \& Scoville, N. 1982, \apj, 258, 467
\bibitem[Zhang et al.(2006)]{zhan06}
Zhang, J.~S., Henkel, C., Kadler, M., et al.\
%Greenhill, L. J.,  Nagar, N.,  Wilson, A.~S., and Braatz, J. A.\
2006, \aap, 450, 933
%
\end{thebibliography}
\end{document}